\documentclass[a4paper,
               ]{jacow}
\usepackage{pdfpages,multirow,ragged2e} %
\makeatletter%
	\ifboolexpr{bool{xetex}}
	 {\renewcommand{\Gin@extensions}{.pdf,%
	                    .png,.jpg,.bmp,.pict,.tif,.psd,.mac,.sga,.tga,.gif,%
	                    .eps,.ps,%
	                    }}{}
\makeatother

\ifboolexpr{bool{xetex} or bool{luatex}} 
 {}                                      
 {\usepackage[utf8]{inputenc}}           

\usepackage[USenglish]{babel}

\ifboolexpr{bool{jacowbiblatex}}%
 {%
  \addbibresource{jacow-test.bib}
  \addbibresource{biblatex-examples.bib}
 }{}
\begin{document}

\title{PIP-II linac Cryogenic Distribution System design challenges}

\author{T. Banaszkiewicz\thanks{tomasz.banaszkiewicz@pwr.edu.pl}, M. Chorowski, P. Duda, M. Stanclik, Wroclaw University of Science and Technology,\\Wroclaw, Poland\\
R. Dhuley, A. Martinez,  W. Soyars, Fermi National Accelerator Laboratory, Batavia, USA }
	
\maketitle

\begin{abstract}
    The PIP-II linac Cryogenic Distribution System (CDS) is characterized by extremely small heat inflows and robust mechanical design. It consists of a Distribution Valve Box (DVB), Intermediate Transfer Line, Tunnel Transfer Line, comprising 25 Bayonet Cans, and ends with a Turnaround Can. Multiple helium streams, each characterized by distinct helium parameters, flow through each of these elements. The CDS geometry allows maintaining an acceptable pressure drop for each helium stream, considering the planned flows and helium parameters in different operation modes. This is particularly crucial for the return line of helium vapors, which return from cryomodules to the cold compressors and thus have very restrictive pressure drop requirements. On both sides of the DVB there are fixed supports for process pipes. One of the DVB design challenges was to route the process pipes in such a way that their shape provided sufficient compensation for thermal shrinkage. This ensures that the forces resulting from thermal shrinkage acting on the cryogenic valves remain at a level acceptable to the manufacturer. The required thermal budget of the CDS was achieved by thermo-mechanical optimization of its components, like process pipes fixed supports in Bayonet Cans.
\end{abstract}

\section{PIP-II CDS overview}
\begin{figure}[!htb]
   \centering
   \includegraphics*[width=1\columnwidth]{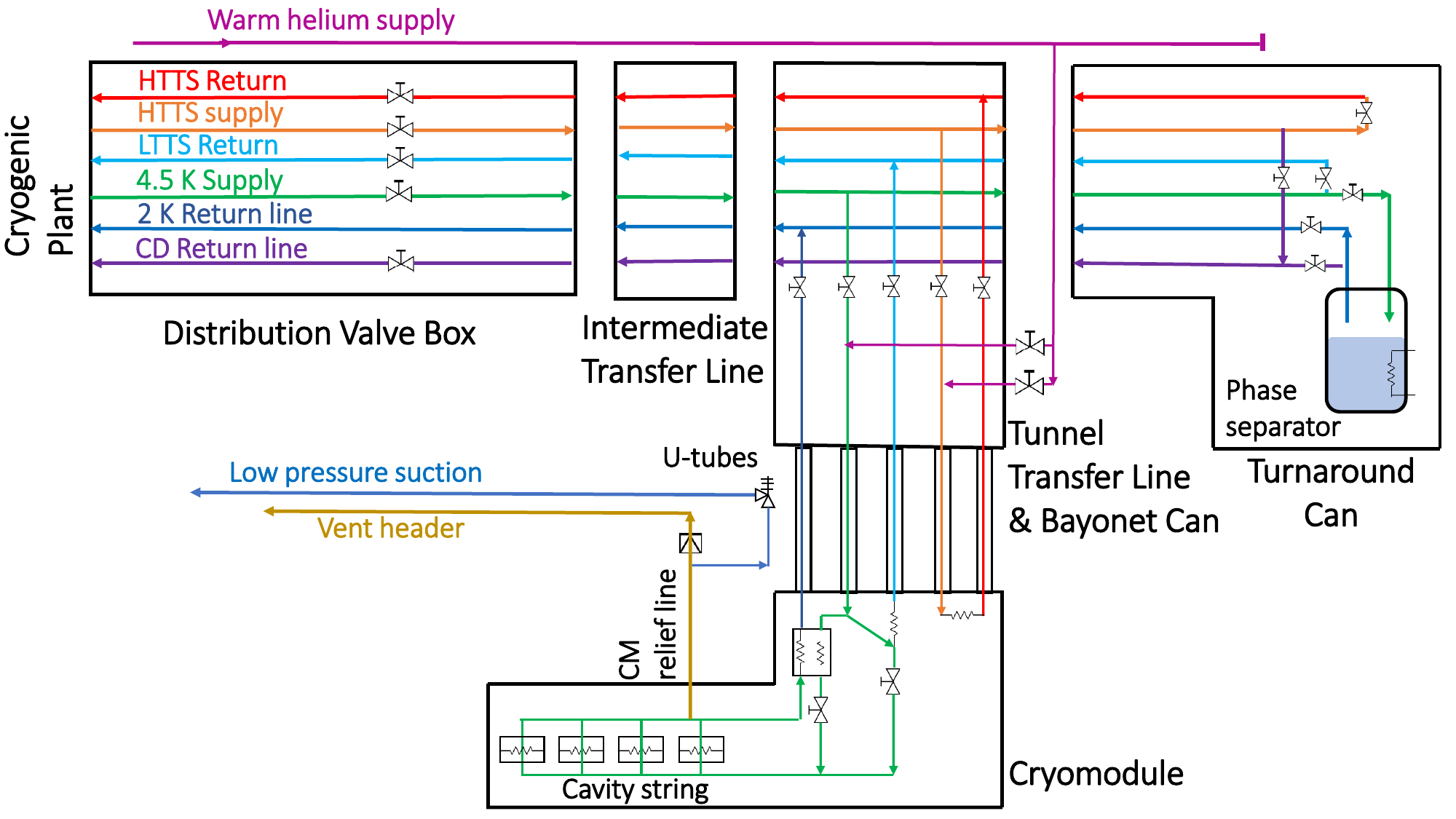}
   \caption{PIP-II CDS layout.}
   \label{fig:CDS}
\end{figure}

The Cryogenic Distribution System is tasked with providing cooling power from the cryogenic plant to the linear accelerator (linac) cryomodules. The simplified layout of the CDS is shown in Fig.~\ref{fig:CDS}
\\The linac consists of five distinct types of cryomodules, each of which is connected to the CDS via its own dismountable, bayonet connection for each process line. The CDS spans approximately 285 meters, running from the Distribution Valve Box at the cryogenic plant interface, through twenty-five Bayonet Cans which are the points where the main helium stream branches into streams that supply the cryomodules, to the Turnaround Can at the linac's end.

The CDS process circuits with size of the main and the branch line and nominal operating conditions (temperatures and pressures) are:
\begin{Itemize}
    \item  \SI{4.5}{K} Supply - DN50/DN15 at \SI{4.5}{K} and \SI{2.9}{bara}
    \item  \SI{2}{K} Return - DN250/DN65 at \SI{3.8}{K} and \SI{3.1}{mbara}
    \item  Low Temperature Thermal Shield Return (LTTS)- DN50/DN15 at \SI{9}{K} and \SI{2.9}{bara}
    \item  High Temperature Thermal Shield Supply (HTTS-S) - DN50/DN20 at \SI{40}{K} and \SI{18}{bara}
    \item  High Temperature Thermal Shield Return (HTTS-R) - DN50/DN20 at \SI{80}{K} and \SI{18}{bara}
    \item  Cooldown Return (CD) - DN80/DN20 at \SI{80}{K}
\end{Itemize}

\section{Pressure drop}
The first discussed challenge was designing the helium transfer line and the helium branched lines in such a way as to meet the constraints of allowable pressure drop budgets. Special attention was given to the helium vapor return line from the cryomodules (\SI{2}{K} Return line). This is a vacuum line used to return the evaporated helium stream from the cryomodules to the cold compressors located in the cryoplant building. During the design of the process lines, all operational modes had to be considered, from the cool-down of the installation through the nominal operating mode to emergency modes. Examples of the helium stream requirements in the nominal operating state and the corresponding pressure drop budget requirements are shown in Tab.~\ref{tab:press_drop_budget}. 
\begin{table}[!hbt]
   \centering
   \caption{CDS process circuit pressure drop budget}
   \begin{tabular}{lcl}
       \toprule
       \textbf{CDS circuit} & \textbf{Press. drop budget} & \textbf{Mass flow} \\
       \midrule
            HTTS                & \SI{280}{mbar} & $\geq$ \SI{65}{g/s}   \\
            \SI{4.5}{K} Supply  & \SI{30}{mbar}  & $\geq$ \SI{172}{g/s}  \\
            LTTS Return         & \SI{30}{mbar}  & $\geq$ \SI{25}{g/s}   \\
            \SI{2}{K} Return    & \SI{4.3}{mbar} & $\geq$ \SI{147}{g/s}  \\
       \bottomrule
   \end{tabular}
   \label{tab:press_drop_budget}
\end{table}
\\The calculations considered all influences on the pressure drop due to the geometry of the individual process circuits, including straight sections, elbows, tees, constrictions, valves, etc. Additionally, geometric constraints related to the actual placement of the individual system elements were taken into account in the calculations. For example, the cryomodules will be placed in a tunnel approximately 14.5~meters below the level of the cryoplant. This makes hydrostatic pressure an important factor, which, depending on the direction of the helium stream flow, can have either a negative or positive impact on the pressure drop.

During the calculations, each circuit was divided into computational segments corresponding to the respective piping elements (straight sections, elbows, valves, etc.) according to the actual geometry. In each segment, the helium stream parameters - temperature, pressure, and mass flow rate - were determined at the inlet to the element. These parameters were used to calculate the pressure drops across the entire CDS. This approach allowed for the determination of helium parameters at every point in the process lines, which positively impacted the system design capabilities and enabled a quick analysis of how changes in geometry would affect the pressure and temperature of helium in the various process pipes.

\section{DVB pipe routing}
\begin{figure}[!htb]
   \centering
   \includegraphics*[width=1\columnwidth]{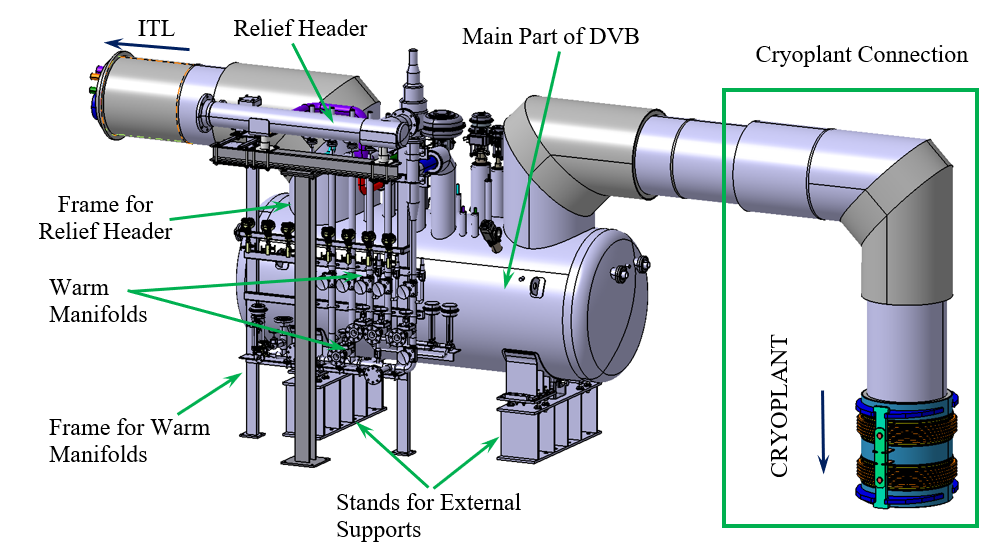}
   \caption{View of the Distribution Valve Box}
   \label{fig:DVB-outer-view}
\end{figure}

The Distribution Valve Box is a component of the CDS that connects the cryoplant with the Intermediate Transfer Line (ITL), which is the multichannel transferline that connects the DVB with the transfer line in the linac tunnel - Tunnel Transfer Line (TTL). The DVB serves several roles. It is a measurement element, housing pressure and temperature sensors for all helium streams. Additionally, coriolis flow meters were placed on the supply lines (\SI{4.5}{K} and HTTS) to measure the mass flow rates of helium fluxes from the cryoplant to supply the entire system. Another purpose of the DVB is to connect the transfer line with auxiliary lines (through the warm manifold). These auxiliary lines enable the system to be purged before the cooling phase. Another function of the DVB is safety. Since the DVB is located at the beginning of the CDS, safety valves have been installed to protect the main cryogenic lines from excessive pressure. A similar solution is used at the end of the CDS in the Turnaround Can. The view of the DVB is shown on Fig.~\ref{fig:DVB-outer-view}.

\begin{figure}[!htb]
   \centering
   \includegraphics*[width=1\columnwidth]{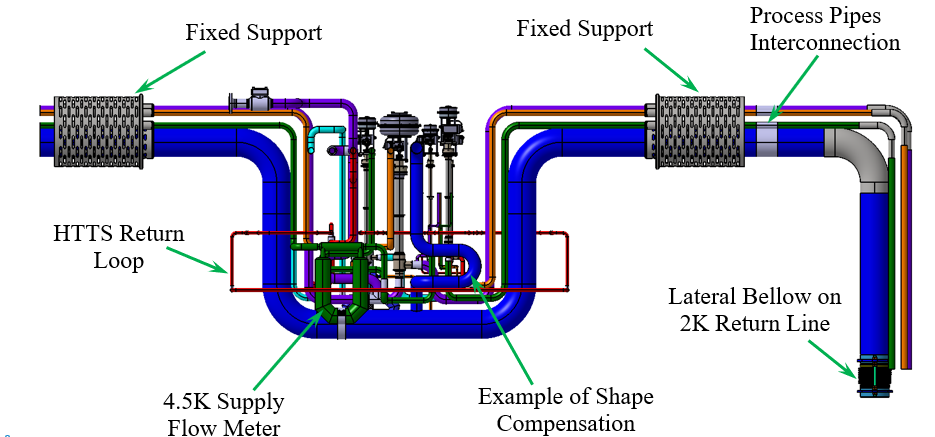}
   \caption{View of the DVB pipes routing}
   \label{fig:DVB-internal}
\end{figure}

For structural reasons, piping fixed supports are located at both ends of the main part of the DVB to support the process pipes (see Fig.~\ref{fig:DVB-internal}). The fixed support is rigidly connecting each process pipe with the vacuum vessel. The support consists of pipe sleeves of different diameters creating heat labyrinth.

Due to the fact that there are fixed supports for the process pipes at both ends of the main part of the DVB, it was necessary to develop a method for compensating thermal shrinkage. The process pipes were routed in such a way that their shape and path would compensate the shrinkage. The design was carried out in such a way that no stresses exceeded the allowable values specified by codes or the requirements of the manufacturers of individual components, such as valves, flow meters, etc. 

This approach allowed for the elimination of bellows and flexible metal hoses, which increase the risk of failure. However, the routing design had to meet more requirements than just compensating for shrinkage inside the process pipes. The DVB is a structure with dimensions limited by its installation location. Consequently, the process pipes had to be routed to fit within a strictly defined space. Within this space, there also had to be room for other components such as supports, thermal shield (TS), and flow meters. Additionally, the DVB design closely collaborated with pressure drop calculations to ensure that the shape needed for shrinkage compensation did not create excessive flow resistance.
The view of the DVB pipe routing is shown on Fig.~\ref{fig:DVB-internal}.

The designed DVB was analyzed using FEM in ANSYS software at each stage of the design process. The calculations were performed taking into account all operational modes, with particular attention to emergency states, which are the primary sources of thermal shocks and can cause sudden stresses. Additionally, the design considered the forces and stresses occurring during the transportation of the device from its manufacturer to the installation site, as well as during the installation itself.

The obtained results indicate that the designed DVB will operate safely in all analyzed load cases. The structural calculation results show that in each load case, the stresses do not exceed the specified allowable values.

The entire Cryogenic Distribution System was divided into segments that allowed for transportation in standard containers from the manufacturing site to the installation site. Similar calculations to those performed for the Distribution Valve Box were conducted for all components of the CDS.

\section{Heat Budget}
The final challenge discussed in this paper was related to heat influx to the cryogenic helium. The entire design of the Cryogenic Distribution System was carried out not only to meet the requirements for pressure drops and allowable forces and stresses on individual components of the system. The CDS was also designed to ensure that the heat influx to each helium stream did not exceed the permissible values specified in the documentation. That includes all the process circuits. The values of the allowable heat load to each helium stream is shown in Tab.~\ref{tab:heat_budget}

\begin{table}[!hbt]
   \centering
   \caption{CDS steady state heat load budgets}
   \begin{tabular}{lrrr}
       \toprule
       \textbf{Process} & \textbf{DVB} & \textbf{ITL} & \textbf{TTL}\\
       \midrule
            HTTS Supply & \SI{9}{W} & \SI{14}{W} & \SI{115}{W}\\
            HTTS Return incl. TS & \SI{123}{W} & \SI{375}{W} & \SI{947}{W}\\
            CD Return  & \SI{5}{W} & \SI{9}{W} & \SI{34}{W}\\
            \SI{4.5}{K} Supply  & \SI{6}{W} & \SI{7}{W} & \SI{32}{W}\\
            LTTS Return  & \SI{6.5}{W} & \SI{6}{W} & \SI{23}{W}\\
            \SI{2}{K} Return & \SI{14}{W} & \SI{36}{W} & \SI{138}{W}\\
       \bottomrule
   \end{tabular}
   \label{tab:heat_budget}
\end{table}

To reduce heat inflow to helium streams, process pipes are placed in vacuum insulation. To minimize heat inflow from radiation, each process pipe will be wrapped with an appropriate number of MLI layers, and each segment of the Cryogenic Distribution System is also equipped with a thermal shield maintained at approximately 80 K by thermal contact with the HTTS Return process line. These elements are standard in Cryogenic Distribution Systems.

The challenge in this project was designing components that connect the process pipes and thermal shield, which are at cryogenic temperatures, to the vacuum jacket, the walls of which remain at ambient temperature~\cite{IOP_CONF}. These components include valve installations, vacuum barriers, and process pipe supports. The supports, in particular, required thorough thermomechanical analysis to ensure sufficient strength and resistance to stability loss due to buckling under the assumed loads, while minimizing heat inflow to the process pipes. The analyses also aimed to ensure that at no point on the external jacket does the temperature drop below the dew point. This is a safeguard against moisture from the air condensing on the cold parts of the cryogenic equipment.

Three general types of internal supports have been developed: the sliding support for the thermal shield, the sliding support for the process pipes, and the fixed support. The sliding support for the thermal shield allows for the centralization of the thermal shield within the vacuum jacket, independent of thermal contractions. The sliding support for the process pipes serves the same purpose but acts as a barrier between the process pipes and the thermal shield. The fixed support is rigidly connecting each process pipe with vacuum vessel. The support consists of pipes of different diameters creating heat labyrinth. Support between warm vacuum vessel and cold process pipes contains intermediate element in shape of a plate. To reduce heat inputs to process pipes the plate is thermally connected with HTTS Return and with thermal shield. For further heat inputs reduction fixed support labyrinth pipes are perforated. Three types of fixed supports were used in the project, which can be termed as weak support, strong support and vacuum barrier.

\begin{figure}[!htb]
   \centering
   \includegraphics*[width=0.8\columnwidth]{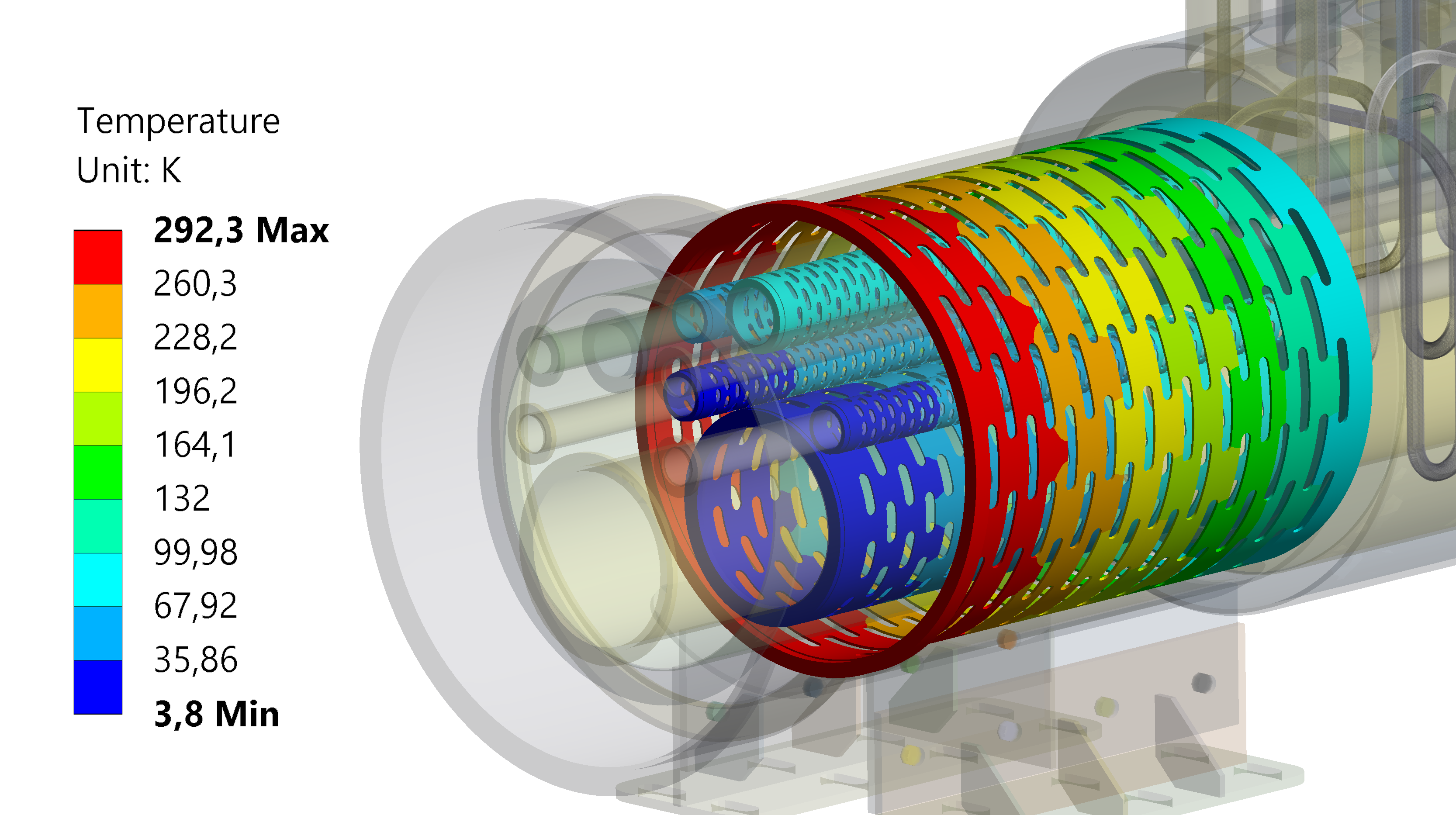}
   \caption{View of the weak fixed support}
   \label{fig:weak-support}
\end{figure}

An example view of the internal weak fixed support is shown in Fig.~\ref{fig:weak-support}

Weak supports are those in which the axial loads acting on them compensate each
other. The strong supports are those in which there are constant non-zero axial loads. The vacuum barriers purpose is to split the TTL vacuum system into smaller sections. The vacuum barrier besides being a barrier is also fixed support for process pipes. The structure of these supports is very similar, the only difference is the thickness of the elements used and the absence of perforations in vacuum barrier.

\flushcolsend
\color{black}
\section{CONCLUSION}
During the design of the Cryogenic Distribution System for the project, numerous challenges were encountered that required active, detailed analysis and a tailored approach. The article presents three of them. Each issue demanded a thorough understanding and a specific solution to ensure the system's effectiveness and reliability. The entire CDS was meticulously designed in accordance with best practices, taking into account the requirements of both the cryoplant and the cryomodules. The design adhered to the codes and to the guidelines set by the manufacturers of the components used in the project. This approach ensured that all aspects of the system met the criteria for performance and safety.

\section{ACKNOWLEDGEMENTS}
The project and the manuscript was created in collaboration between Wroclaw University of Science and Technology and Fermi Research Alliance, LLC under Contract No. DE-AC02-07CH11359 with the U.S. Department of Energy, Office of Science, Office of High Energy Physics.

%
%
\ifboolexpr{bool{jacowbiblatex}}%
	{\printbibliography}%
	{%
	
	
} 
%
%


\end{document}